\documentclass[aps, prb, reprint, superscriptaddress]{revtex4-2}

\usepackage{graphicx}
\usepackage{dcolumn}
\usepackage{bm}
\usepackage{upgreek}
\usepackage{color,soul}
\usepackage{amsmath}

\begin{document}

\title{Reservoir computing in a lithium-based magneto-ionic device
}

\author{Sreeveni Das}
\affiliation{NanoSpin, Department of Applied Physics, Aalto University School of Science, P.O. Box 15100, FI-00076 Aalto, Finland}
\author{Rhodri Mansell}
\email[Corresponding author: ] {rhodri.mansell@aalto.fi}
\affiliation{NanoSpin, Department of Applied Physics, Aalto University School of Science, P.O. Box 15100, FI-00076 Aalto, Finland}
\author{Aarne Piha}
\affiliation{NanoSpin, Department of Applied Physics, Aalto University School of Science, P.O. Box 15100, FI-00076 Aalto, Finland}
\author{Luk{\'{a}}{\v{s}} Flaj{\v{s}}man}
\affiliation{NanoSpin, Department of Applied Physics, Aalto University School of Science, P.O. Box 15100, FI-00076 Aalto, Finland}
\author{Maria-Andromachi Syskaki}
\affiliation{Singulus Technologies AG, 63796 Kahl am Main, Germany}
\author{J\"{u}rgen Langer}
\affiliation{Singulus Technologies AG, 63796 Kahl am Main, Germany}
\author{Sebastiaan van Dijken}
\email[Corresponding author: ]{sebastiaan.van.dijken@aalto.fi}
\affiliation{NanoSpin, Department of Applied Physics, Aalto University School of Science, P.O. Box 15100, FI-00076 Aalto, Finland}

\begin{abstract}
In-materio computing exploits the intrinsic physical dynamics of materials to perform complex computations, enabling low-power, real-time data processing by embedding computation directly within physical layers. Here, we demonstrate a voltage-controlled magneto-ionic device that functions as a reservoir computer capable of forecasting chaotic time series. The device consists of a crossbar structure with a Ta/CoFeB/Ta/MgO/Ta bottom electrode and a LiPON/Pt top electrode. A chaotic Mackey-Glass time series is encoded into a voltage signal applied to the device, while 2D Fourier transforms of voltage-dependent magnetic domain patterns form the output. Performance is influenced by the input rate, smoothing of the output, the number of elements in the reservoir state vector, and the training duration. We identify two distinct computational regimes: short-term prediction is optimized using smoothed, low-dimensional states with minimal training, whereas prediction around the Mackey-Glass delay time benefits from unsmoothed, high-dimensional states and extended training. Reservoir computing metrics reveal that slower input rates are more tolerant to output smoothing, while faster input rates degrade both memory capacity and nonlinear processing. These findings demonstrate the potential of magneto-ionic systems for neuromorphic computing and offer design principles for tuning performance in response to input signal characteristics.
\end{abstract}

\maketitle
\section{Introduction}

Physical reservoir computing is a promising neuromorphic computing paradigm that uses the intrinsic dynamics of physical systems to perform temporal information processing and pattern recognition tasks \cite{LUKOSEVICIUS2009127,Tanaka2019,Yan2024}. Unlike recurrent neural networks, which require extensive training of internal weights, reservoir computing simplifies the process by training only the output weights. Various physical implementations have been demonstrated using optical \cite{Duport2012,VanderSande2017}, optoelectronic \cite{Paquot2012,Liu2022}, mechanical \cite{Fernando2003pattern,Dion2018}, electronic \cite{Du2017,Moon2019,Zhong2021}, and magnetic systems \cite{Torrejon2017,Bourianoff2018,Watt2020,Allwood2023,Lee2024,Everschor-Sitte2024}, each exploiting distinct temporal dynamics and offering prospects for high-speed, energy-efficient computation \cite{Yan2024}.

Magneto-ionic devices, systems that modulate magnetic properties through voltage-driven ion migration, hold significant promise for neuromorphic computing. These devices naturally exhibit short-term plasticity, mimicking the behavior of biological synapses \cite{Mishra2019,Monalisha2023,Monalisha2024,Bernard2025,Das2025}. Furthermore, the interplay between ionic transport and magnetization dynamics gives rise to nonlinear responses and memory effects, making magneto-ionic systems well-suited for reservoir computing \cite{Monalisha2024,Das2025}. 

Magneto-ionic effects have been demonstrated across a wide range of material systems involving various mobile ions, including oxygen \cite{Bi2014,Bauer2015}, hydrogen \cite{Tan2018,Huang2021}, nitrogen \cite{deRojas2020,Rojas2022}, and lithium \cite{DasGupta2014,Ameziane2022,Ameziane2023a}. Reversible voltage-induced ion migration enables dynamic control over key magnetic properties, such as magnetic moment \cite{DasGupta2014,deRojas2020,Rojas2022}, magnetic anisotropy \cite{Bi2014,Bauer2015,Tan2018,Ameziane2022}, the Dzyaloshinskii-Moriya interaction \cite{Srivastava2018}, and the Ruderman–Kittel–Kasuya–Yosida (RKKY) interaction \cite{Kossak_RKKY_Voltage_2023,Ameziane2023a}. Low-voltage magneto-ionic actuation has also been employed to reversibly control skyrmion nucleation and annihilation \cite{Fillion2022,Ameziane2023}.

\begin{figure}[htbp]
    \centering
    \includegraphics[width=1.0\linewidth]{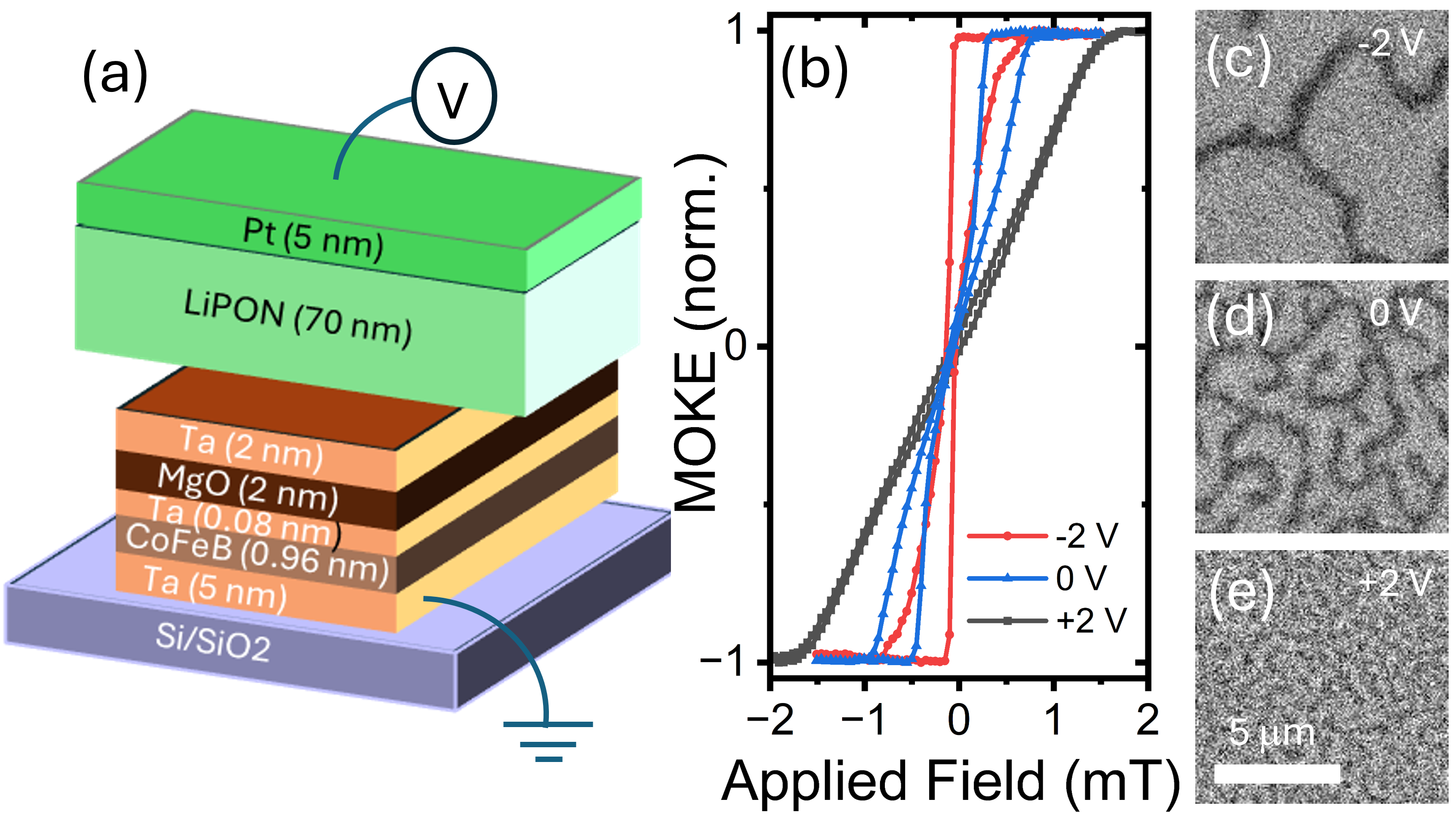}
    \caption{(a) Schematic of the magneto-ionic device, comprising a crossbar junction with a layered Ta/CoFeB/Ta/MgO/Ta bottom electrode and LiPON/Pt top electrode. Application of a positive voltage to the top electrode drives Li$^+$ ions from the LiPON layer toward the bottom electrode, enabling voltage-controlled modulation of the magnetic domain state in the CoFeB layer. (b) Out-of-plane magnetic hysteresis loops measured by MOKE microscopy under different applied voltages, illustrating voltage-induced changes in magnetic behavior. (c)-(e) MOKE microscopy images showing the magnetic domain structure in the CoFeB layer at a fixed out-of-plane magnetic field of 0.3 mT under applied voltages of (c) -2 V, (d) 0 V, and (e) +2 V.}
    \label{fig1}
\end{figure}

\begin{figure*}[bhtb]
    \centering
    \includegraphics[width=1.0\linewidth]{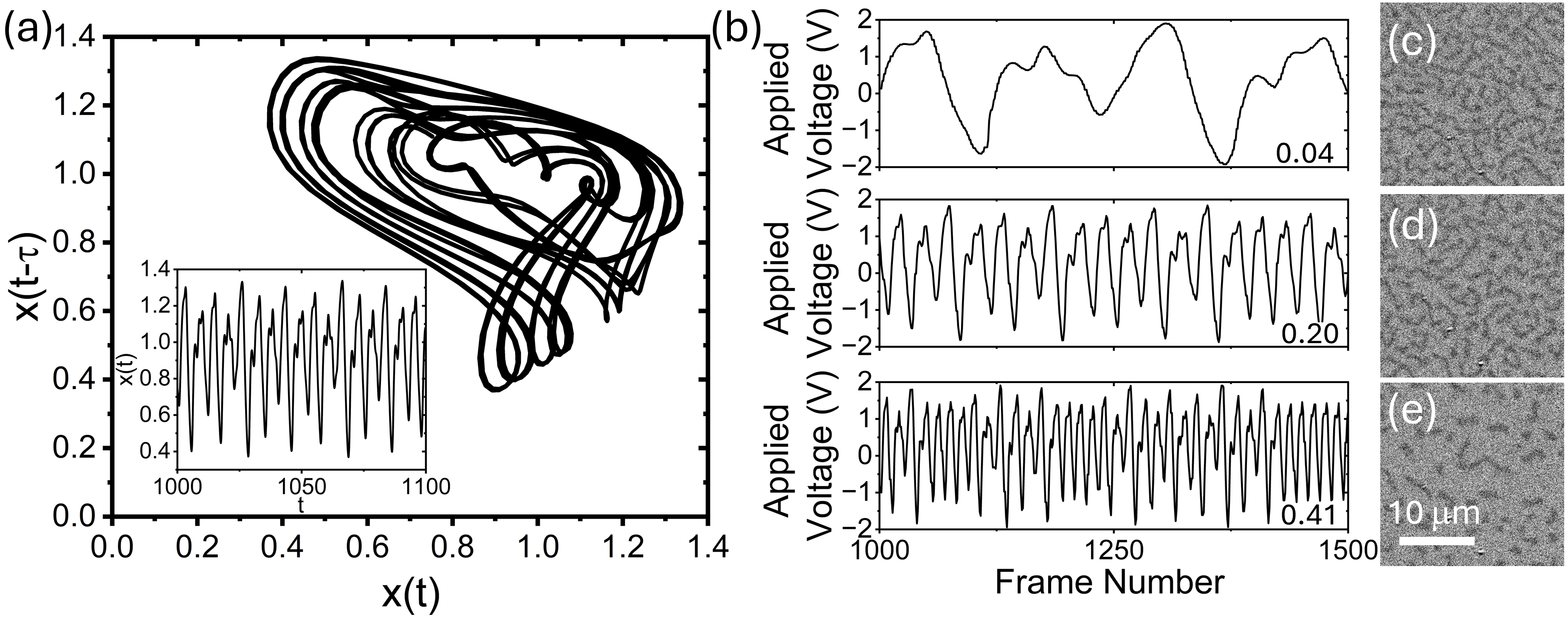}
    \caption{(a) Delay-embedded trajectory of the Mackey-Glass signal used in this study. The inset shows the same 100-step segment of the signal as a time series. (b) Representative voltage waveforms used to drive the magneto-ionic reservoir, obtained by linearly mapping the Mackey-Glass signal onto a voltage range of -2 V to +2 V. The input rates shown are 0.04 (top), 0.20 (middle), and 0.41 (bottom) Mackey-Glass steps per MOKE microscopy frame. (c)-(e) Zoom-ins of MOKE microscopy images captured at the (c) 1000th, (d) 1050th, and (e) 1100th frame in a video recorded at an input rate of 0.20 Mackey-Glass steps per frame.}
    \label{fig2}
\end{figure*}

Device architectures inspired by solid-state batteries and supercapacitors, such as LiCoO$_2$/LiPON/Co and LiPON/CoFeB heterostructures \cite{Ameziane2022,Ameziane2023,Monalisha2024}, have enabled reversible modulation of magnetic properties via lithium ion migration. Lithium insertion into the magnetic layer induces significant changes in magnetic anisotropy, leading to pronounced modifications of the magnetic domain structure. These devices also demonstrate excellent voltage cycling endurance, exceeding 10$^5$ cycles \cite{Ameziane2023}, underscoring their potential for use in reservoir computing applications. 

The Mackey-Glass (MG) equation is widely used as a benchmark to evaluate the temporal processing capabilities of reservoir computing systems \cite{Soriano2015,Lee2024,Sun2023}. Originally developed to model delayed physiological feedback, this delay differential equation generates chaotic behavior for specific parameter regimes \cite{MackeyGlass1977}.  The form used here is:
\begin{equation}
    \frac{\textrm{d}x}{\textrm{d}t} =  \frac{ax(t-\tau)}{1+x(t-\tau)^n} -bx(t),
\end{equation}
where $x$ is the Mackey-Glass signal, $t$ is time, and $a,b, \tau$ and $n$ are control parameters. 

In this work, we implement a reservoir computing system based on a magneto-ionic device, in which voltage-driven ion migration induces the nonlinear dynamics required for temporal information processing. The system's performance is evaluated using a Mackey-Glass time-series prediction task, serving as a representative benchmark for chaotic signal processing. We show how input rate, state-vector dimensionality, and causal smoothing modify prediction ability, yielding task-adaptive operating points without hardware changes.

\section{Results and Discussion}

\begin{figure*}[htbp]
    \centering
    \includegraphics[width=1.0\linewidth]{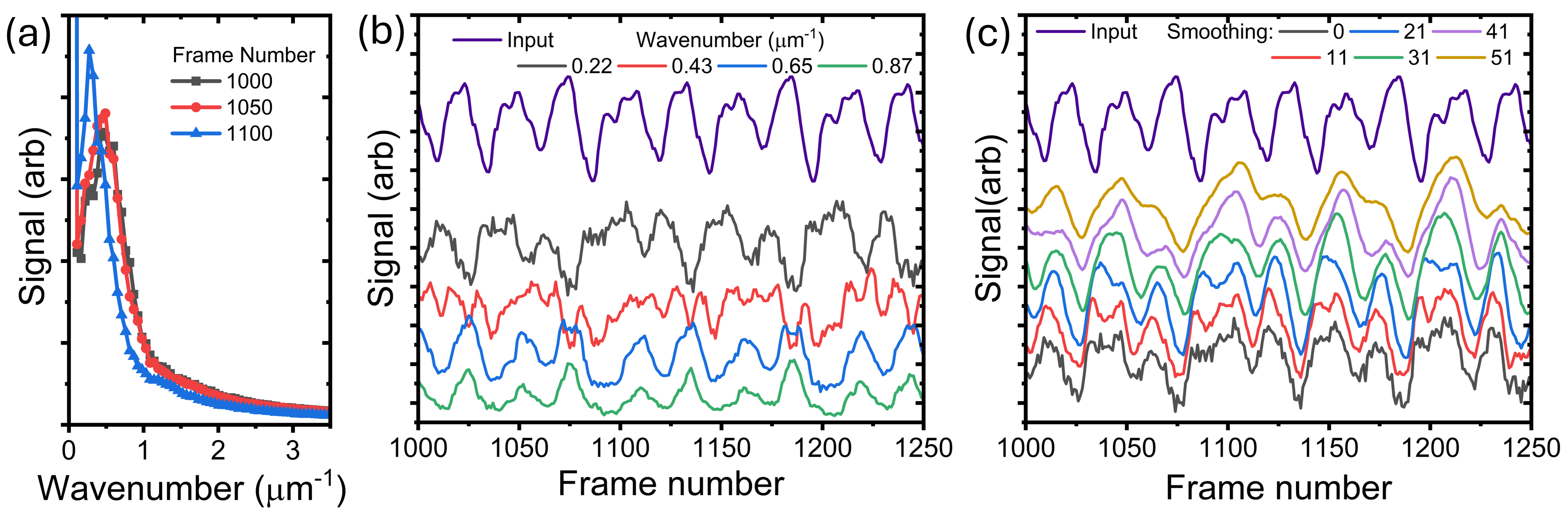}
    \caption{(a) Radially averaged 2D Fourier transforms of the full-size versions of the MOKE microscopy images from Fig.\ 2, recorded at an input rate of 0.20 Mackey-Glass steps per frame. (b) Temporal evolution of the reservoir signal for selected wavenumber components as a function of video frame number, recorded at an input rate of 0.20 Mackey-Glass steps per frame. The corresponding input signal is shown for comparison. (c) Effect of causal Savitzky-Golay filtering on the reservoir states for an input rate of 0.20 Mackey-Glass steps per frame. Reservoir responses are shown for second-order polynomial smoothing with window lengths of 0, 11, 21, 31, 41, and 51 data points at 0.22 $\upmu$m$^{-1}$. The input signal is again shown for reference.}
    \label{fig3}
\end{figure*}

Figure 1(a) illustrates the device structure, consisting of a 110 $\upmu$m $\times$ 80 $\upmu$m crossbar junction with a layered bottom electrode of Ta(5) / CoFeB(0.96) / Ta(0.08) / MgO(2) / Ta(2), and a top electrode of LiPON(70) / Pt(5), where all thicknesses are in nanometers (see Methods for fabrication details). The CoFeB layer exhibits perpendicular magnetic anisotropy (PMA). The lithium phosphorus oxynitride (LiPON) solid-state electrolyte enables Li$^+$ ion transport under an applied bias between the Pt top electrode and the metallic bottom layers. Applied positive voltages cause Li$^+$ ions to migrate towards the bottom electrode. Li ions likely intercalate into the MgO/Ta layers, which has been shown to reduce the interfacial PMA in similar devices \cite{Ameziane2022,Ameziane2023,Das2025}. Conversely, applying negative voltages extracts Li$^+$ ions and leads to an increased PMA. Figure 1(b) shows magnetic hysteresis loops measured via magneto-optical Kerr effect (MOKE) microscopy under an out-of-plane magnetic field for different applied voltages. At an applied -2 V, the loop exhibits increased squareness, indicating enhanced interfacial PMA. At 0 V, the loop becomes more tilted, reflecting a reduction in anisotropy. Under +2 V, the loop is significantly slanted with minimal hysteresis, consistent with anisotropy suppression due to Li$^+$ ion intercalation \cite{Ameziane2022}. Figures\ 1(c)-1(e) present MOKE microscopy images captured at a constant out-of-plane magnetic field of 0.3 mT under various bias conditions. At -2 V (Fig.\ 1(c)), a few broad stripe domains are visible. At 0 V (Fig.\ 1(d)), the domain density increases and individual stripes become narrower. At +2 V (Fig.\ 1(e)), the domain pattern becomes much denser and finer, indicating a progressive reduction in magnetic anisotropy with increasing positive voltage.

Figure 2 illustrates the generation of the Mackey-Glass signal and its injection into the magneto-ionic reservoir. For the parameters in Eq.\ 1, we used $a = 2$, $b = 1$, $\tau = 2$, and $n = 9.65$, which produces a chaotic time series \cite{Glass1979}. The delay-embedded trajectory of this signal is shown in Fig.\ 2(a), revealing a characteristic orbit around a chaotic attractor. The inset displays a 100-step segment of the time series, highlighting irregular oscillations with intermittent amplitude modulations. In this paper we will refer to a unit increase in the time, $t$ in the Mackey-Glass equation as a Mackey-Glass step. Figure 2(b) shows the voltage waveforms used to drive the reservoir, obtained by linearly mapping the Mackey-Glass signal onto a voltage range of -2 V to +2 V. The three panels correspond to three of the ten input rates used in the experiments. These rates are defined relative to the 16 frames-per-second (fps) acquisition rate of the MOKE microscopy system. In the top panel, the slowest rate is shown, corresponding to 0.04 Mackey-Glass steps per MOKE microscopy frame. The middle panel represents an intermediate rate of 0.20 steps per frame, and the bottom panel shows the fastest rate used, 0.41 steps per frame. These voltage waveforms are applied to the magneto-ionic device under an out-of-plane magnetic bias field of 0.3 mT, while MOKE microscopy videos of 672 $\times$ 512 pixels are recorded at 16 fps, each consisting of approximately 2400 frames (see Methods for details). Figures\ 2(c)-2(e) show zoom-ins of MOKE microscopy frames captured at the slowest input rate, corresponding to the 1000th, 1050th, and 1100th Mackey-Glass input steps. The application of voltage leads to distinct changes in the domain state, altering the density of stripe domains and skyrmions. The stripe domains and skyrmions have different energy barriers for creation and annihilation, where the energy barriers are also dependent on the applied voltage as well as the surrounding domain pattern \cite{Monalisha2024,Mansell2023,Das2025}. This leads to a complex time evolution of the magnetic state under a varying applied voltage as required for reservoir computing.

\begin{figure*}[bhtp]
    \centering
    \includegraphics[width=1.0\linewidth]{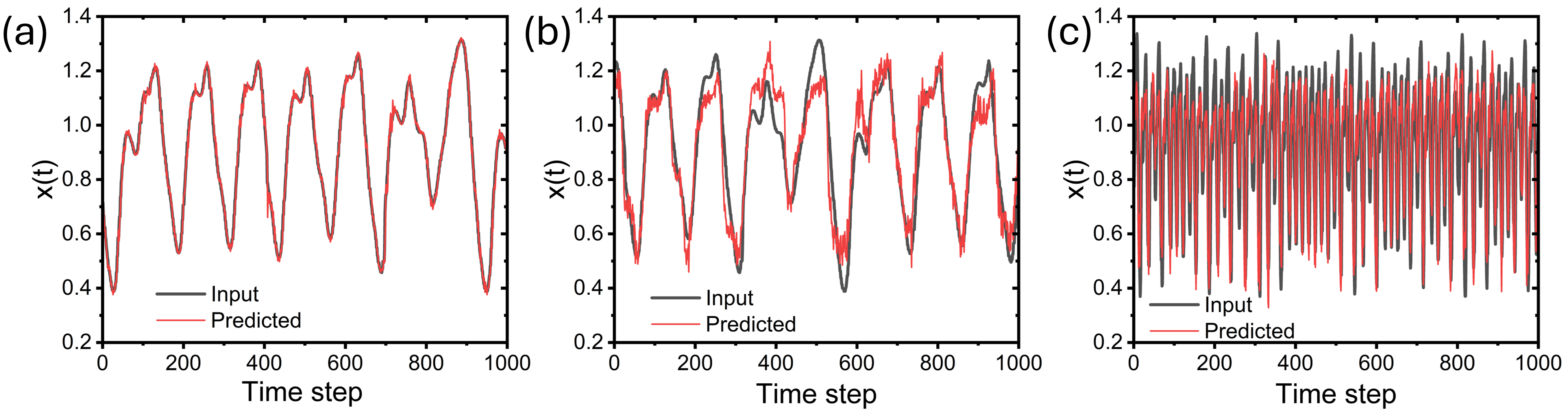}
    \caption{Predictions of the Mackey-Glass signal under varying input and model parameters. The time step is referenced to the video frame rate. (a) One-step ahead prediction at an input rate of 0.04 Mackey-Glass steps per frame, using the full 64-component reservoir state vector, 20 training frames, and a 51-point Savitzky-Golay filter. (b) 50-step-ahead prediction at the same input rate, performed with a reduced 16-component state vector, 400 training frames, and no smoothing. (c) Five-step-ahead prediction at a higher input rate of 0.33 Mackey-Glass steps per frame, using a 64-component state vector, 1000 training frames, and no smoothing.}
    \label{fig4}
\end{figure*}

After recording the video sequence for each input rate, the frames were cropped to 450 $\times$ 450 pixels, binarized and subjected to a spatial 2D Fourier transform. The resulting spectra were radially averaged around the $k = 0$ wave vector to obtain the reservoir signal as a function of wavenumber (see also the Methods). Figure\ 3(a) shows three such traces corresponding to the domain images shown in Fig.\ 2(c)-(e). Each Fourier spectrum is truncated to a 1D vector comprising at most 64 components, which form the reservoir state vector at each timestep. Variations in magnetic domain shape, size and density, driven by the Mackey-Glass voltage inputs, lead to distinct changes in these output vectors. Figure\ 3(b) illustrates the temporal evolution of selected components of the reservoir state vector, alongside the original Mackey-Glass input signal. Due to the nonlinear mapping from input voltage to magnetic domain pattern, followed by spatial Fourier processing, the reservoir components exhibit dynamics that are highly transformed relative to the input. To examine the impact of noise reduction on signal integrity, Fig.\ 3(c) analyzes the effect of a causal Savitzky–Golay filter applied to the reservoir states in the time domain. Second-order polynomial smoothing was applied using window lengths of 0, 11, 21, 31, 41, and 51 data points, with each window ending at the current time point. This analysis highlights the trade-off between noise suppression and signal fidelity: moderate smoothing (11–21 points) effectively reduces experimental noise while preserving key temporal features, whereas heavier smoothing (41–51 points) introduces noticeable signal distortion. 

Once the reservoir state vectors are generated from the video frames, the reservoir computer is trained (see Methods). To evaluate the effect of altering the state vector, the system is trained across different configurations. The size of the state vector is varied from by including Fourier components up to a cutoff wavenumber, always starting from the lowest wavenumber values and including all values up to the maximum. This gives a state vector varying from 4 to 64 values. Smoothing is systematically varied from no filtering to a 51-point Savitzky-Golay filter, as described above. For each configuration, the training length is also varied between 20 and 1000 frames. Finally, all combinations are evaluated across ten input rates, with forward prediction lengths ranging from 1 to 100 time steps, where the time step refers to that of the video frames rather than the Mackey-Glass time.

\begin{figure*}[htbp]
    \centering
    \includegraphics[width=0.8\linewidth]{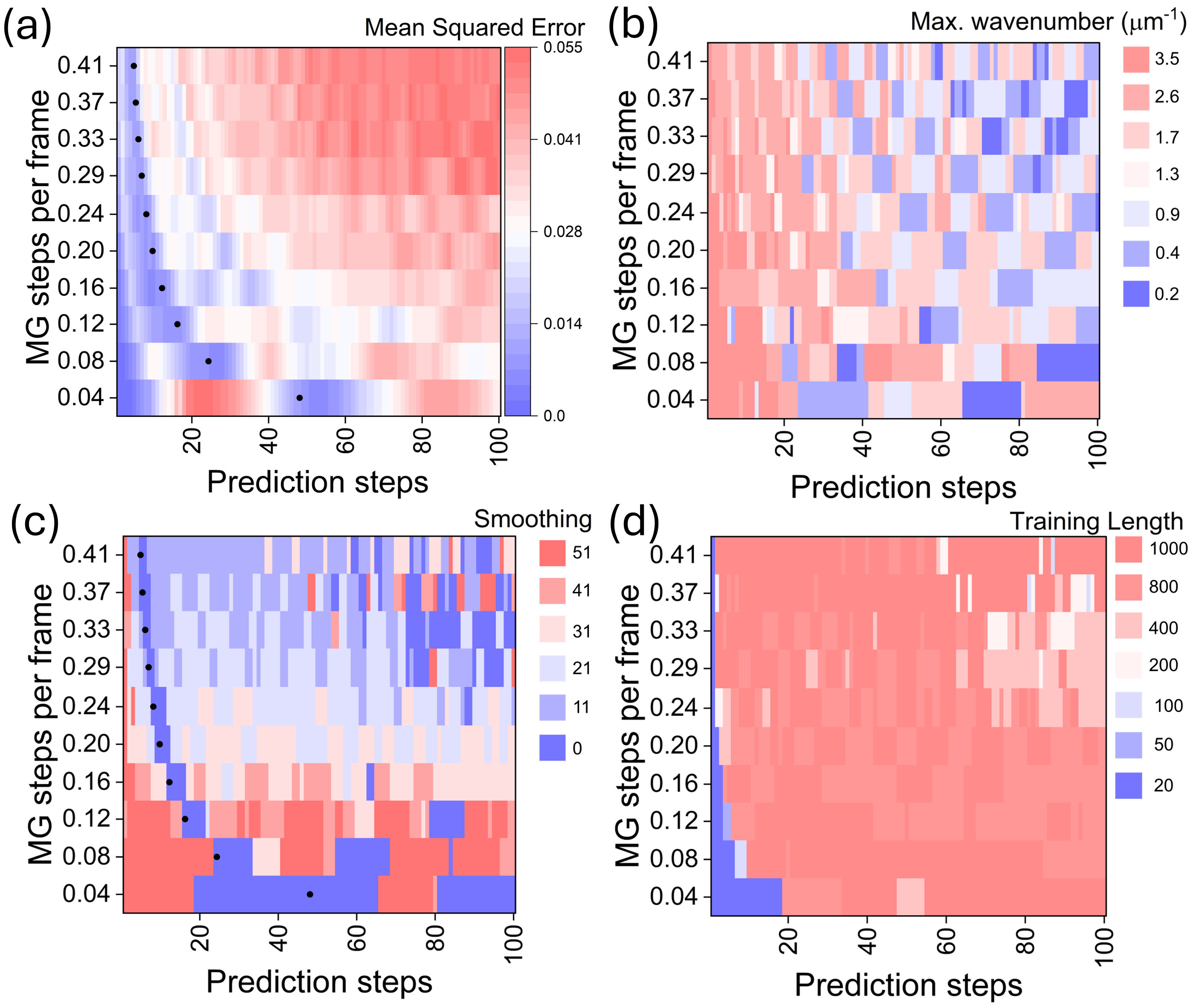}
    \caption{Optimization results for predicting the Mackey-Glass signal across various prediction horizons and input rates. The prediction step is referenced to the video frames. (a) Minimum mean squared error (MSE) achieved for each combination of prediction horizon and input rate, determined by scanning over all combinations of state vector sizes, training lengths, and smoothing parameters. The black dots mark the number of steps in the reference time of the video frames equal to $\tau$. (b) Optimal maximum wavenumber component (i.e., number of wave vector components) that minimizes the prediction error for each setting. (c) Optimal Savitzky-Golay smoothing window length yielding the lowest prediction error across all conditions. The black dots mark the number of steps in the reference time of the video frames equal to $\tau$. (d) Optimal number of training frames resulting in the smallest prediction error for each prediction horizon and input rate.}
    \label{fig5}
\end{figure*}

Figure 4 presents three representative prediction tasks, each illustrating the combination of input and model parameters that yielded the best performance, as measured by the mean squared error (MSE). Each data point represents an average over 1000 trials, where the prediction task is repeated with a one-frame forward shift to build statistical confidence. Figure 4(a) shows a one-step-ahead prediction at the slowest input rate (0.04 Mackey-Glass steps per frame), demonstrating excellent agreement with the target signal. This optimal result was achieved using the full 64-component state vector, minimal training (20 frames), and heavy smoothing (51-point Savitzky-Golay filter). Figure 4(b) shows a 50-step-ahead prediction at the same input rate. This prediction horizon corresponds approximately to the Mackey-Glass delay time ($\tau$), a regime where the system's temporal correlations are strongest. As expected, the prediction accuracy is lower than for single-step forecasting. The best performance in this case was obtained with a reduced state vector (16 components, wavenumbers up to 0.9 $\upmu$m$^{-1}$), moderate training length (400 frames), and no smoothing. Notably, prediction accuracy is higher during quasi-periodic segments of the signal, while larger deviations occur in more chaotic regions. Figure 4(c) shows five-step-ahead prediction for a faster input rate (0.33 Mackey-Glass steps per frame), optimized using a 64-component state vector, extensive training (1000 frames), and no smoothing. A systematic underestimation of high-amplitude input values is observed. This effect is attributed to magnetic saturation: at large positive voltages, the system enters a dense stripe-domain configuration, reducing the dynamic range of the reservoir and limiting prediction accuracy in this regime. 

Figure\ 5 summarizes the optimal reservoir parameters across a range of input rates and prediction horizons. Figure\ 5(a) shows the minimum MSE obtained from all combinations of maximum wavenumber, training lengths, and smoothing parameters. The error landscape reveals two distinct regimes of enhanced predictive performance: (i) a short-term forecasting region at a low number of prediction steps, dominated by linear extrapolation, and (ii) a broader minimum centered around the number of prediction steps equivalent to the  Mackey-Glass delay time $\tau$ (marked with a black dot), with a weaker secondary minimum near $2\tau$. For example, at an input rate of 0.08 Mackey-Glass steps per frame, these regions appear around 1-5 prediction steps, again at 25 steps (approximately $\tau$ for this case), and faintly around 50 steps. Figure\ 5(b) shows the optimal maximum wavenumber corresponding to the minimum error. While larger wavenumbers generally yield better predictions, deviations from this trend occur in regions with poor overall performance. In these cases, using a state vector with fewer elements may help prevent overfitting and improve generalization. Figure\ 5(c) presents the optimal smoothing window length. Two distinct patterns emerge: increased smoothing is favored at slower input rates, where noise dominates and signal distortion is minimal. In contrast, near the delay time $\tau$, zero smoothing is consistently preferred, highlighting the importance of preserving high-frequency components for accurate prediction in this regime. Figure 5(d) displays the optimal training length. Longer training generally enhances performance across most conditions. However, a notable exception occurs at short prediction horizons, where minimal training leads to better results, likely due to reduced overfitting in this linear prediction regime.

\begin{figure*}[htbp]
    \centering
    \includegraphics[width=1.0\linewidth]{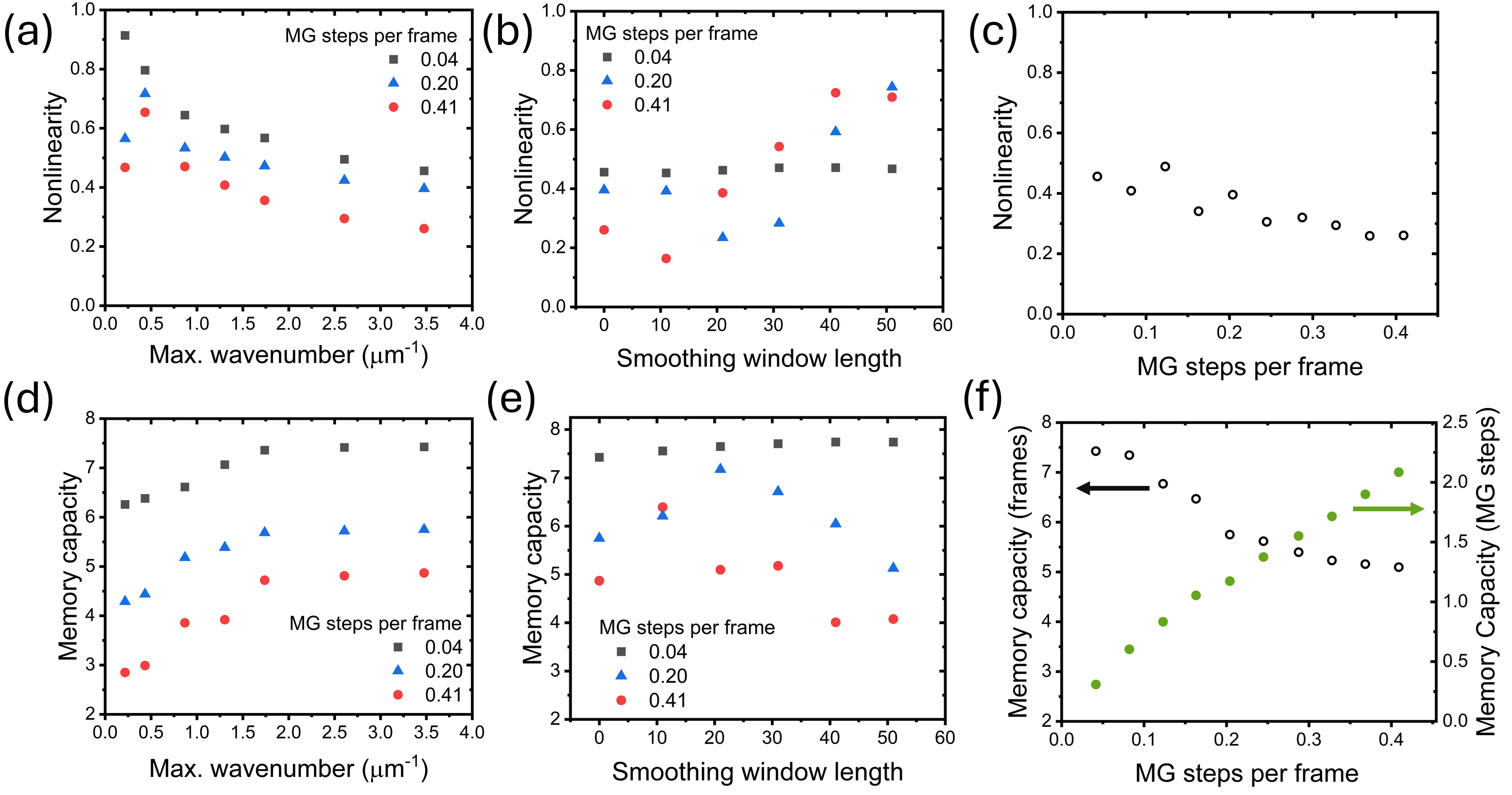}
    \caption{Nonlinearity and memory capacity of the magneto-ionic reservoir under varying input and processing conditions. (a) Nonlinearity as a function of the maximum wavenumber used to create the state vector, measured at input rates of 0.04, 0.20, and 0.41 Mackey-Glass steps per frame, with no smoothing applied. (b) Nonlinearity as a function of Savitzky-Golay smoothing window length, evaluated for the same input rates using  the full 64-component state vector (i.e. wavenumbers up to 3.5 $\upmu$m$^{-1}$). (c) Nonlinearity as a function of input rate, obtained with a 64-component state vector and no smoothing. (d) Memory capacity as a function of the maximum wavenumber used to create the state vector, evaluated at input rates of 0.04, 0.20, and 0.41 Mackey-Glass steps per frame without smoothing. (e) Memory capacity as a function of smoothing window length, using a 64-component state vector at the same three input rates. (f) Memory capacity as a function of input rate, expressed in terms of both the number of video frames (left axis) and the equivalent number of Mackey-Glass steps (right axis). These results are obtained using a 64-component state vector and no smoothing.}
    \label{fig6}
\end{figure*}

Figure 6 evaluates the reservoir's response using two key metrics: nonlinearity and memory capacity (see Methods for details). Figure 6(a) plots the nonlinearity metric versus the maximum wavenumber (at zero smoothing) for three input rates (0.04, 0.20, and 0.41 Mackey-Glass steps per frame). Nonlinearity peaks for small maximum wavenumbers and declines beyond that point. This reflects that the most nonlinear transformations of the input occur near the peak of the radially averaged Fourier transform, around 0.5 $\upmu $$m^{-1}$ (see Fig.\ 3(a)). At larger wavenumbers, the reservoir outputs become more homogeneous, resulting in reduced nonlinearity. Figure 6(b) shows how smoothing affects the nonlinearity metric for a 64-component state vector. At the slowest input rate, nonlinearity remains relatively unaffected by smoothing, indicating that the relationship between input and output states is preserved. At higher input rates, however, a non-monotonic behavior emerges: moderate smoothing reduces nonlinearity, but excessive smoothing significantly distorts the reservoir states, leading to an artificial increase in nonlinearity. Figure\ 6(c) illustrates the effect of input rate on nonlinearity for the case of 64 state vector components and zero smoothing. Increasing the input rate leads to a reduction in nonlinearity, likely because faster voltage variations do not allow sufficient time for complex, thermally activated domain configuration to evolve before the input changes again.

The lower panels of Fig.\ 6 examine memory capacity, using the same methodology as for the nonlinearity. As shown in Fig.\ 6(d), memory capacity increases with state vector size and eventually saturates. Slower input rates yield higher capacity, suggesting more persistent temporal correlations in the magnetic response. Figure\ 6(e) shows the impact of smoothing: for slow input rates, smoothing has minimal effect. For faster rates, moderate smoothing enhances memory capacity (as it implicitly incorporates past information), but extensive smoothing degrades the performance due to signal distortion. Figure\ 6(f) explores the impact of input rate on memory capacity. When measured relative to the video frame rate (left axis), memory capacity decreases with increasing input rate. However, when the memory capacity is considered in terms of the number of Mackey-Glass steps (right axis) instead of the number of frames, memory capacity increases with input rate. This highlights how the apparent memory of the reservoir depends on the timescale used for evaluation. Taken together, the metrics data confirm a fundamental trade-off between nonlinearity and memory capacity, consistent with prior observations in reservoir computing systems \cite{Love2023}: mechanisms that enhance one often diminish the other. 

In summary, our results demonstrate that the magneto-ionic device effectively maps temporal voltage inputs onto high-dimensional magnetic states through lithium ion migration, enabling accurate forecasting of chaotic dynamics in the Mackey-Glass system. Optimal predictive performance depends critically on the forecasting horizon and system parameters. For short-term predictions, optimal results are achieved using smoothed reservoir states, minimal training, and large state vectors, which enhance linear extrapolation and reduced noise. In contrast, long-term predictions, especially near the Mackey-Glass delay time $\tau$, benefit from unsmoothed data, extensive training, and maximal reservoir dimensionality to capture nonlinear temporal dynamics. Notably, prediction accuracy peaks around $\tau$ and $2\tau$, consistent with known recurrence times in chaotic systems. High voltages were found to drive the system into saturated, densely packed stripe-domain configurations, where small voltage changes yielded minimal magnetic response, thereby reducing sensitivity and predictive accuracy. Overall, these findings underscore the potential of magneto-ionic systems for reservoir computing and highlight the need to tailor reservoir characteristics, including nonlinearity, memory, and readout strategy, to the specific demands of the forecasting task. With appropriate tuning, these systems offer a powerful platform for complex time-series prediction, even in highly nonlinear and chaotic regimes.

\section*{Methods}
\subsection*{Fabrication}
The Ta(5) / CoFeB(0.96) / Ta(0.08) / MgO(2) / Ta(2) stack was grown via magnetron sputtering at room temperature in an Ar atmosphere using a Singulus Rotaris sputtering system at Singulus Technologies AG. The film was patterned to form the bottom electrode using photolithography and Ar ion-beam milling. A second photolithography step was used to define the cross-bar top electrode. The top electrode, consisting of LiPON (70 nm) / Pt (5 nm), was grown by magnetron sputtering using a Kurt J. Lesker system at Aalto University. The LiPON layer was formed by sputtering a Li$_3$PO$_4$ target in a N$_2$ atmosphere. The junction was defined using a lift-off process giving a junction size of 110 $\times$ 80 $\upmu$m.

\subsection*{Data collection}
Images and videos were acquired using an Evico magneto-optical Kerr effect (MOKE) microscope in the polar configuration. A 100$\times$ objective lens and a 1.6$\times$ projection lens were used to image the magneto-ionic junction. Voltage signals were applied using a Digilent Analog Discovery 2 device. The Mackey-Glass signal was generated in MATLAB, rescaled to a $\pm$2 V range, and applied to the magneto-ionic device at various rates using the Digilent waveform generator. MOKE videos were recorded at 16 frames per second, with approximately 2400 frames captured per input rate.

\subsection*{Image processing}
The original 672$\times$512 pixel images were cropped to 450$\times$450 pixels to remove areas affected by a variation in the focus, and processed in MATLAB using a sequence of functions: imdiffusefilt for noise reduction, imadjust for contrast enhancement, and imbinarize for binarization. 2D fast Fourier transforms were then applied to the binary images, and the resulting spectra were radially averaged around the $k = 0$ wave vector, with a binning width equal to two pixels of the fast Fourier transform to reduce noise. These spectra were truncated into 1D vectors with 64 components, which served as the reservoir state vectors. To further reduce noise, a second-order causal Savitzky–Golay filter was applied, with each smoothed point computed using only preceding data points.

\subsection*{Training}

Using the output of the 2D fast Fourier transforms a reservoir state matrix, S, is created with dimensions [training length $\times$ state vector]. An output vector D, of size [training length] was created from the Mackey-Glass equation stepped forward from the input by the prediction step length. Training was performed using ridge regression to give weight matrices, $W_\mathrm{out}$:
\begin{equation}
    W_\mathrm{out} = (S^TS + \alpha I)^{-1}S^TD,
\end{equation}
where $I$ is the identity matrix, and alpha is a small constant. The training length was varied from 20 to 1000. The number of columns in the state matrix was varied from 4 to 64 components, created by taking the radially-averaged Fourier transform wavenumber starting from the k$=0$ wavenumber up to a varying maximum value. The prediction step was varied from 1 to 100, and each configuration was repeated 1000 times by advancing the training and prediction window forward by one frame. The weight matrices were then applied to unseen data to make the prediction. The final prediction error was calculated as the average mean squared error of the predicted versus expected output across the 1000 repetitions. For each input rate and prediction horizon, the optimal combination of state vector size, smoothing window, and training length was identified based on the minimum error.

\subsection*{Metrics}
Metrics were calculated following the approach by Lee \textit{et al}. \cite{Lee2024}. The nonlinearity metric quantifies the fraction of the reservoir output that cannot be explained by a linear mapping from past inputs. For each reservoir output component $y_s$, a linear model is fitted using eight lagged input values:
\begin{equation}
    \hat y_s = a_0 + \sum_{i=1}^8 a_i x(t-i)
\end{equation}
where $\hat y_s$ is the predicted output, $x(t-i)$ are lagged inputs, and the coefficients $a_i$ are obtained via linear regression. The predictive quality of the linear model is evaluated using the coefficient of determination:
\begin{equation}
    \textrm{R}^2_s\left[\hat y_s{,y_s} \right] = 1 -\frac{\sum_{i=1}^n (y_{s,i}-\hat y_{s,i})^2}{\sum_{i=1}^n(y_{s,i}-\bar y_s)^2},
\end{equation}
where $\bar y_s = \frac{1}{n}\sum_{i=1}^n y_{s,i}$, and $n$ is the number of time steps in the test dataset. The nonlinearity is then defined as: 
\begin{equation}
    \textrm{NL} = 1 - \textrm{mean}(\textrm{R}^2_s\left[\hat y_s{,y_s} \right]), 
\end{equation}
where the mean is taken over all relevant components of the reservoir state vector. A higher nonlinearity indicates a greater deviation from linear behavior in the reservoir's response to the input signal.

The short-term memory capacity quantifies the ability of the reservoir to reconstruct past inputs based solely on its current output state. It is evaluated by applying a linear model to predict past input values from the reservoir output: 
\begin{equation}
    \hat x(t-i) = \sum_{s=1}^n w_sy_s(t),
\end{equation}
where $\hat x(t-i)$ is the predicted input at lag $i$, $y_s(t)$ are the state vector components at time $t$, and $w_s$ are regression coefficients obtained via linear least squares fitting. The reconstruction quality for each lag is quantified using the coefficient of determination $R^2$, and the overall memory capacity (MC) is defined as:
\begin{equation}
    \textrm{MC} = \sum_{i=1}^8 \textrm{R}^2\left[\hat x(t-i),x(t-i)\right].
\end{equation}
This metric captures how much information about recent inputs is retained and linearly accessible from the reservoir's internal state. A higher memory capacity indicates better short-term recall performance.

\section*{Acknowledgments}

This project received funding from the European Union’s Horizon 2020 research and innovation program under the Marie Skłodowska-Curie Grant Agreement No. 860060 “Magnetism and the effects of Electric Field” (MagnEFi). It was also supported by the Research Council of Finland (Grant No. 338748). We acknowledge the provision of facilities by Aalto University at OtaNano-Micronova Nanofabrication Centre, and the computational resources provided by the Aalto Science-IT project.

\bibliography{MGRef}

\end{document}